\documentclass[reprint,superscriptaddress,amsmath,amssymb,longbibliography,aps,prb]{revtex4-2}
\usepackage{hyperref}
\usepackage[version=4]{mhchem}
\usepackage{graphicx}
\usepackage{amsmath}
\usepackage{color}
\usepackage{soul}
\usepackage{pdfpages} 
\usepackage{pgffor} 
\usepackage[printfigures]{figcaps} 
\figcapsoff 

\newcommand{\srrho}{Sr$_{2}$RhO$_{4}$}

\newcommand{\down}{\downarrow}
\newcommand{\up}{\uparrow}

\pdfpageattr {/Group << /S /Transparency /I true /CS /DeviceRGB>>}
\def\supplementfilename{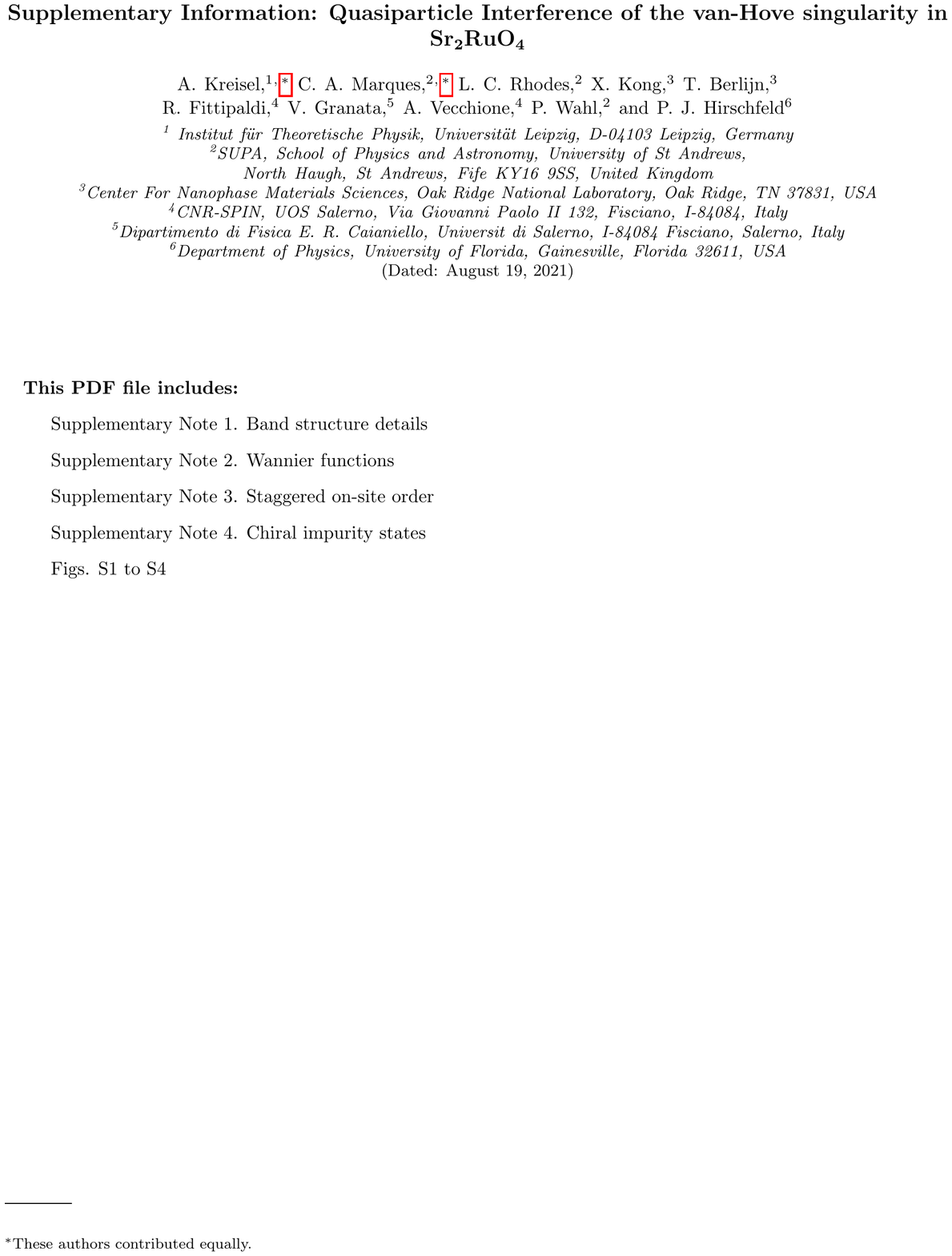}

\pdfximage{\supplementfilename}
\def\numbersupplementpages{\the\pdflastximagepages}

\newif\ifarXiv
\arXivtrue 
\makeatletter
\AtBeginDocument{\let\LS@rot\@undefined}
\makeatother
 
\begin{document}


\title{Quasiparticle Interference of the van-Hove singularity in \ce{Sr2RuO4}}

\author{A. Kreisel}
\thanks{These authors contributed equally.}
\affiliation{
Institut f\" ur Theoretische Physik, Universit\"at Leipzig, D-04103 Leipzig, Germany
}

\author{C. A. Marques}
\thanks{These authors contributed equally.}
\affiliation{%
SUPA, School of Physics and Astronomy, University of St Andrews, North Haugh, St Andrews, Fife KY16 9SS, United Kingdom
}%

\author{L. C. Rhodes}
\affiliation{%
SUPA, School of Physics and Astronomy, University of St Andrews, North Haugh, St Andrews, Fife KY16 9SS, United Kingdom
}%

\author{X. Kong}
\affiliation{Center For Nanophase Materials Sciences, Oak Ridge National Laboratory, Oak Ridge, TN 37831, USA}
\author{T. Berlijn}
\affiliation{Center For Nanophase Materials Sciences, Oak Ridge National Laboratory, Oak Ridge, TN 37831, USA}
\author{R. Fittipaldi}
\affiliation{CNR-SPIN, UOS Salerno, Via Giovanni Paolo II 132, Fisciano, I-84084, Italy}
\author{V. Granata}
\affiliation{Dipartimento di Fisica “E. R. Caianiello”, Universit\`a di Salerno, I-84084 Fisciano, Salerno, Italy}
\author{A. Vecchione}
\affiliation{CNR-SPIN, UOS Salerno, Via Giovanni Paolo II 132, Fisciano, I-84084, Italy}
\author{P. Wahl}
\affiliation{%
SUPA, School of Physics and Astronomy, University of St Andrews, North Haugh, St Andrews, Fife KY16 9SS, United Kingdom
}%
\author{P. J. Hirschfeld}
\affiliation{Department of Physics, University of Florida, Gainesville, Florida 32611, USA}

\date{\today}

\begin{abstract}
The single-layered ruthenate \ce{Sr2RuO4} is one of the most enigmatic unconventional superconductors.  While for many years it was thought to be the best candidate for a chiral $p$-wave superconducting ground state, desirable for topological quantum computations, recent experiments suggest a singlet state, ruling out the original $p$-wave scenario. The superconductivity as well as the properties of the multi-layered compounds of the ruthenate perovskites are strongly influenced
by a van Hove singularity in proximity of the Fermi energy. Tiny structural distortions move the van Hove singularity across the Fermi energy with dramatic consequences for the physical properties. Here, we determine the electronic structure of the van Hove singularity in the surface layer of \ce{Sr2RuO4} by quasiparticle interference imaging. We trace its dispersion  and demonstrate from a model calculation accounting for the full vacuum overlap of the wave functions that its detection is facilitated through the octahedral rotations in the surface layer.
\end{abstract}


\maketitle

\section{Introduction}
Strontium Ruthenate, \ce{Sr2RuO4}, has played a leading role in discussions of unconventional superconductivity since its discovery almost three decades ago\cite{maeno_superconductivity_1994,Mackenziereview,Sigrist2005,Maenoreview,Kallin2012,Mackenzie2017}. Much of the interest in the community centered on the possibility of chiral $p$-wave pairing, but the compound has also attracted attention simply because of its structural similarity to the cuprates, Fermi liquid behavior at low temperatures, and the availability of very clean samples with high quality surfaces. 
Recently, several new experimental results\cite{Pustogow19,ghosh2020thermodynamic,benhabib2020jump,grinenko2020split} have called into question the NMR results
on which the traditional triplet pairing scenario was based\cite{Ishida98_wrong}, providing evidence for spin-singlet rather than triplet pairing and leading to a renaissance in the quest to identify the exact pairing state of this fascinating material.
In principle, direct measurement of the superconducting gap by, e.g., Angular Resolved Photoemission Spectroscopy (ARPES), could provide important guidance, as it did in the cuprates. 
However the energy scales involved, such as the transition temperature of only 1.5K in \ce{Sr2RuO4}, or the temperature at which the metamagnetic transitions in \ce{Sr3Ru2O7} occur, are beyond the capabilities of current ARPES instruments. STM is a more appropriate tool, which due to its very high energy resolution that can be achieved at low temperatures and the ability to obtain information about the momentum- and phase-resolved structure of the superconducting gap through quasiparticle interference (QPI) imaging\cite{hoffman_imaging_2002,hanaguri_coherence_2009} promises to resolve the most pressing questions about the superconducting properties of \ce{Sr2RuO4}.

Significant progress has recently been made towards achieving this goal.
The electronic structure in the bulk of \ce{Sr2RuO4} near the Fermi energy is well-known to consist of weakly hybridized 1D sheets ($\alpha$ and $\beta$) of $d_{xz/yz}$ character, as well as a 2D $d_{xy}$ sheet ($\gamma$) that hybridizes with both. The $\gamma$-band has a van-Hove singularity which in the bulk is $\sim 14\,\mathrm{meV}$ above the Fermi energy\cite{shen_evolution_2007}, but whose energy depends sensitively on small structural changes \cite{Sunko2019} with significant consequences for the superconductivity\cite{steppke_strong_2017}. This van Hove singularity plays not only an important role in the properties of \ce{Sr2RuO4}, but also of the bilayer and trilayer ruthenates, where the van-Hove singularity has been suggested to be the origin of the metamagnetic behaviour\cite{binz_metamagnetism_2004}. 

\begin{figure*}
    \begin{center}
    \includegraphics[width=\linewidth]{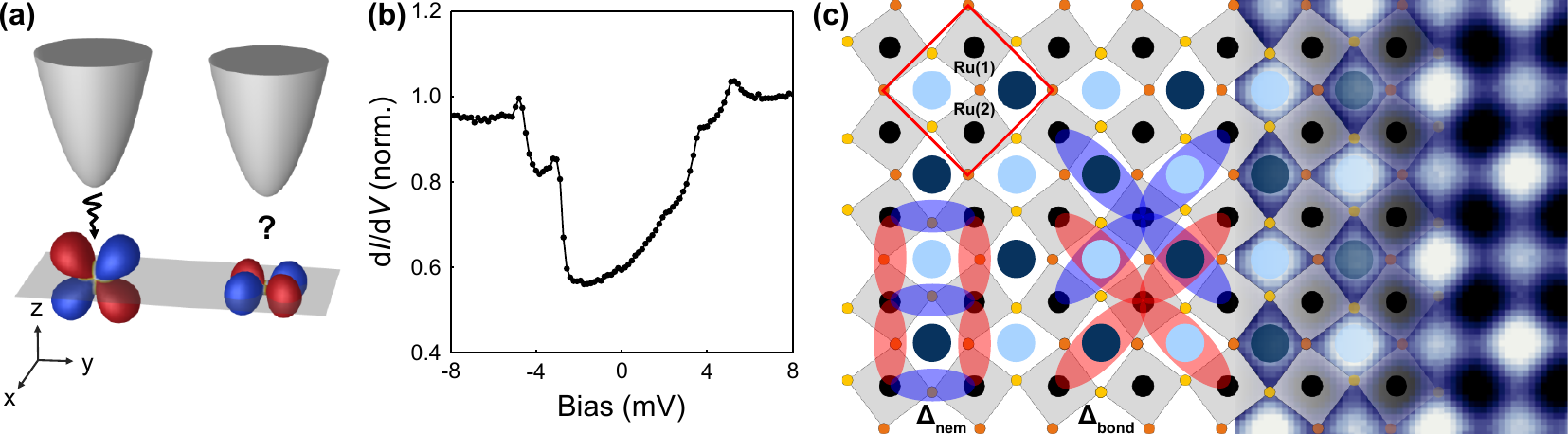}
    \end{center}
    \caption{\textbf{Electron tunneling and nematicity at the surface of \ce{Sr2RuO4}.} (a) The out-of-plane symmetry of a $d_{yz}$ orbital suggests a high probability of electron tunneling from a metallic tip and significant contribution to the tunneling current. For an in-plane atomic orbital, such as $d_{xy}$, the overlap with the tip orbitals is small, suggesting negligible contribution. (b) High-resolution differential conductance (d$I$/d$V$) spectrum at $T=76\mathrm{mK}$, showing the characteristic gap-like structure at the surface of \ce{Sr2RuO4} ($V_\mathrm{set} =8 \mathrm{mV}$, $I_\mathrm{set}=500.2 \mathrm{pA}$, $V_\mathrm{L} =155 \mu\mathrm{V}$). Four peaks can be clearly identified which are associated with the $d_{xy}$-derived $\gamma$-band\cite{Marques2020}. (c) Sketch of the structure of the reconstructed surface of \ce{Sr2RuO4} with RuO$_6$ octahedra rotated by the angle $\vartheta=6^\circ$. The experimentally observed checkerboard charge order (right hand side) is equivalent to the breaking of $C_4$-symmetry rendering the two Sr atoms (dark blue and light blue) and oxygen bonds along the horizontal and vertical direction inequivalent (indicated by orange and yellow circles)\cite{Marques2020}. This symmetry breaking is described by a nematic order with $d_{x^2-y^2}$ symmetry ($\Delta_\mathrm{nem}$, red and blue indicate positive and negative sign, respectively, of the order parameter). The checkerboard charge order can be accounted for either through an additional staggered bond-centred order ($\Delta_\mathrm{bond}$, colors as for the nematic order parameter), or through a staggered on-site order with different on-site energies for the $d_{xy}$ band at Ru(1) and Ru(2).}
    \label{fig1}
\end{figure*}

In STM measurements, the situation has been less clear, and not all bands found in the bulk have been detected so far: Firmo et al.\cite{Firmo2013} argued that
the gap observed in tunneling corresponds to that on the 1D $d_{xz}/d_{yz}$ bands,
with the $d_{xy}$ band not contributing to tunneling spectra but still exhibiting a sizeable gap due to proximity coupling. Tunneling to the STM tip is argued to happen primarily due to the $d_{xz}/d_{yz}$ states in the sample, with the justification that $d_{xy}$ states associated with the $\gamma$-band have lobes that lie in the plane, while $d_{xz}/d_{yz}$ states have lobes pointing out of the surface plane towards the tip (see also Fig.~\ref{fig1}(a)).
Similarly, in recent QPI experiments in the normal state of \ce{Sr2RuO4}, the observed patterns were attributed to bands with $d_{xz}$ and $d_{yz}$ character\cite{Wang2017}. The expected Bogoliubov-QPI in \ce{Sr2RuO4} for a chiral order parameter has been
previously investigated theoretically within a lattice Green's function framework that neglected
the surface reconstruction and effect of the tunneling matrix elements\cite{Akbari2013}. A recent attempt to characterize the momentum-space structure of the superconducting gap was made by Sharma et al.\cite{Sharma20}, who propose that their data was consistent with $d$-wave pairing and the signal from QPI due to the $d_{xz}$/$d_{yz}$ bands. 

These results thus all raise the question what the role of the $\gamma$ sheet is which has escaped detection in STM experiments so far. Detection of the $\gamma$ band will allow one to to decide how large the gap is on this sheet, and understand whether it arises only from coupling to the $\alpha$ and $\beta$ bands as argued in Ref.~\onlinecite{Firmo2013}, or whether the STM tip is simply insensitive to states of $d_{xy}$-character\cite{Wang2017, Sharma20}. Recent STM experiments report signatures of the $d_{xy}$ band in tunneling spectra \cite{Marques2020} (Fig.~\ref{fig1}(b)) and an apparent absence of the superconducting gap, raising important questions about its role in superconductivity.
Understanding and reconciling these seemingly contradictory interpretations is of primary importance in the effort to understand the QPI in this material and ultimately determine the momentum space structure of the superconducting  gap in \ce{Sr2RuO4}.

An important aspect of studies of clean surfaces of \ce{Sr2RuO4} is the surface reconstruction, which arises as a spontaneous rotation of the RuO$_6$ octahedra similar to the crystal structure in the bulk of \ce{Sr3Ru2O7}\cite{matzdorf_ferromagnetism_2000}.  This reconstruction has recently been shown to influence QPI patterns on \ce{Sr2RuO4}, and is possibly relevant for the low-energy electronic structure\cite{Wang2017,Marques2020}. 

Here we resolve the mystery of the seeming absence of the $\gamma$-band from tunneling spectroscopy through combination of a phenomenological model of the low energy electronic structure with ab-initio calculations of the tunneling matrix elements and ultra-low temperature scanning tunneling microscopy experiments. Our results demonstrate QPI of the  $\gamma$-band and settle the normal state electronic structure and QPI in the surface layer of \ce{Sr2RuO4} and thus provide a reference for modelling of the superconducting QPI. 
We show that for the unreconstructed surface tunneling into $d_{xy}$ states is suppressed primarily due to the alternating character of the Bloch $d_{xy}$ function at momenta near the van Hove point, and to a lesser extent due to the weaker extension of this Wannier function in the $z$ direction. In the case of the reconstructed surface, we find that the amplitude of the $d_{xy}$ tunneling is enhanced due to an admixture of $d_{z^2}$ orbital character mediated by the rotation of the oxygen octahedra. 

\section{Results}
\subsection{Open questions - coupling to $d_{xy}$ states and emergent orders}
As discussed in the introduction, we begin with the premise that understanding the QPI in the normal state \cite{Wang2017,Sharma20,Marques2020} will be essential to identification of the symmetry and structure of the superconducting gap in \ce{Sr2RuO4} by STM.  There are several features of the measured patterns that are challenging to interpret.  The first is the dramatic suppression of the features in the measured spectrum that originate from bands with dominant $d_{xy}$ orbital content, compared to  a  calculation of $N(\bf q,\omega)$ using the lattice Green's function with  bulk electronic bands as done in Refs. \cite{Wang2017,Sharma20}, where QPI was modelled by ignoring any $d_{xy}$ contribution to the trace and hence density of states.
This suppression of scattering features associated with the $\gamma$-band has been discussed in terms of $d_{xy}$ orbitals coupling weakly to the tip due to the location of their lobes in the $xy$ plane\cite{Firmo2013}, or to orbital-selective decoherence of these orbitals\cite{RomerPRL}.  As a practical matter, QPI data has often been analyzed simply ignoring $d_{xy}$ contributions \cite{Wang2017}, which seems to work up to a point.  Nevertheless, there are $\bf q$-peaks observed in these patterns that correspond to scattering from Fermi surface points close to the van Hove singularities dominated by $d_{xy}$ states\cite{Marques2020}, so a complete theory needs to account for these features as well.  
 
A second set of puzzles is associated with the checkerboard charge order and associated nematicity of the \ce{Sr2RuO4} surface\cite{Marques2020}. The phenomena associated with these include a chirality of impurity states emanating from defects on different  sublattices, and the bias dependence of the intensity of the atomic peaks.  While these have been definitively associated with the reconstructed surface, their origin is unclear.  Here we show that all these phenomena can be explained in a natural way with a combination of two types of coexisting orders, a nematic order and a staggered bond order, and by accounting for the vacuum tail of the involved electronic states. 

\subsection{Tight-binding model}
We start from a model constructed from Wannier functions of the three Ru orbitals $d_{xz},d_{yz}$ and $d_{xy}$ which have been established as the relevant electronic states in the normal state Fermi surface of \ce{Sr2RuO4} by ARPES experiments~\cite{Tamai2019,Veenstra2014,Haverkort08,Zabolotnyy13}.
The surface Wannier states and the corresponding tight binding model are obtained from an ab-initio calculation (see Methods section
for additional information).
The lattice Hamiltonian is given by
\begin{equation}
 H_0=\sum_{\mathbf{R},\mathbf{R}'}\sum_{\alpha\beta} t_{\mathbf{R},\mathbf{R}'}^{\alpha \beta} c_{\mathbf{R},\alpha}^\dagger c_{\mathbf{R}',\beta}^{\phantom{\dagger}},
 \label{H_0}
\end{equation}
where  $t_{\mathbf{R},\mathbf{R}'}^{\alpha \beta}$ are hopping elements between the elementary cells described by the vectors $\mathbf{R}$ and $\mathbf{R}'$. See Supplementary Note 1 and Supplementary Fig.~1 for a plot of the band structure and Supplementary Note 2 and Supplementary Fig.~2 for a discussion of the Wannier functions.  $\alpha$ and $\beta$ are combined orbital and spin indices, the chemical potential enters as an on-site term and spin-orbit coupling has been added as (complex-valued) onsite terms to represent $H_{\mathrm{SOC}}=\lambda \mathbf{L}\cdot \mathbf{S}$ in the usual way, see Methods section
for details.

\subsection{Nematicity and checkerboard charge order}
Unlike theoretical models for bulk \ce{Sr2RuO4}, the surface reconstruction requires adoption of an elementary cell with two Ru atoms to describe the surface electronic structure as observed in STM and ARPES experiments. Therefore, we work in a basis that contains 6 Wannier states per spin, three of them centered at Ru(1) and three at Ru(2), see Fig.~\ref{fig1}(c). As suggested by the crystal structure of the related material \ce{Sr3Ru2O7}, in the surface layer the oxygen octahedra around Ru(1) atoms rotate anticlockwise and clockwise around Ru(2) (details on the resulting crystallographic structure are given in Appendix \ref{app_dft}),
giving rise to the reconstructed surface and yielding qualitatively different tunneling properties as we discuss later.
To fully describe the low energy electronic structure, two order parameters are required: a nematic term and a term describing the checkerboard charge order observed experimentally.
Such terms could in principle be understood from microscopic models as instabilities of the electronic structure as worked out for the related material \ce{Sr3Ru2O7}\cite{Raghu2009,Lee2009,Puetter2010} and do modify hopping amplitudes of the $d_{xy}$ orbital.
The nematic term affects the nearest neighbor (NN) hopping between Ru atoms. To describe the checkerboard charge order, we introduce a staggered bond order on the next nearest neighbor (NNN) hopping terms between Ru atoms, see Fig.~\ref{fig1}(c). 
The symmetry properties of the two orders are identical once the
octahedral rotation is present in the reconstructed surface layer\cite{Marques2020}.
The explicit form of the Bloch Hamiltonian as $6\times 6$ matrix is not convenient to discuss at this point since the two sublattice basis already breaks the $C_4$ symmetry. We therefore discuss the symmetries in real space: The NN nematic term is of $d_{x^2-y^2}$ symmetry (see Appendix \ref{app_nem})
and induces a positive shift of the hopping amplitude along [0,1] (red ellipse connecting NN Ru atoms) and a negative shift of the hopping amplitude along [1,0] (blue ellipses) of amplitude $\pm \Delta_{\mathrm{nem}}$,  preserving the translational symmetry between the Ru(1) and Ru(2) atoms. 
 
\begin{figure*}
  \begin{center}
    \includegraphics[width=\linewidth]{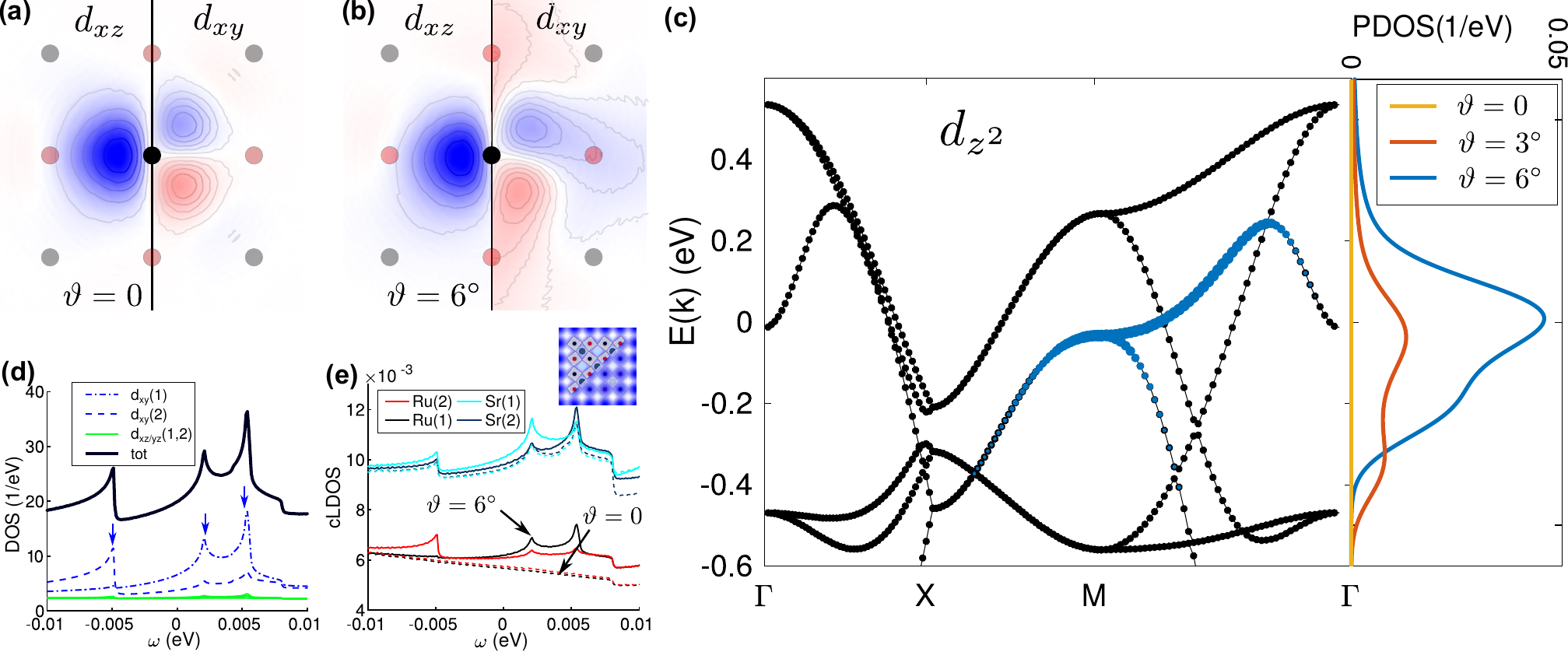}
    \end{center}
    \caption{\textbf{Tunneling probabilities and the surface reconstruction.} (a)  Calculated cross sections in the $xy$ plane 5\AA{} above the unreconstructed \ce{Sr2RuO4} surface of $d_{xz}$ and $d_{xy}$ Wannier functions. (b) Same, but for reconstructed surface in the presence of $\vartheta=6^\circ$ O-octahedron rotation. Black dot represents the position of the Ru(1) atom, gray dots the neighboring Ru(1) atoms and light red dots the Ru(2) atoms. (c) DFT band structure for the reconstructed surface with $\vartheta=6^\circ$ employed in this work, with $d_{z^2}$ orbital weight highlighted in blue. The right panel shows the projected $d_{z^2}$ density of states for $\vartheta=6^\circ$ as in the panel on the left, as well as for $\vartheta=3^\circ$ and $0^\circ$ (red, yellow). (d) Lattice density of states from model described in text (black: total; blue: $d_{xy}$ on the Ru(1) and Ru(2) atoms exhibiting peaks at the positions marked with arrows, green: $d_{xz/yz}$ (approximately fourfold degenerate) with only tiny structure within the energy range shown)  (e) Continuum density of states evaluated 5\AA{} above the surface above the two inequivalent Ru atoms and above the two inequivalent Sr atoms, see inset. Solid lines show result using Wannier functions with octahedral rotation and exhibit features at the peaks of the $d_{xy}$ lattice DOS, while a calculation using Wannier functions without octahedral rotation yields a completely flat cLDOS above the Ru positions within this energy range.}
    \label{Fig:Wannier}
\end{figure*}

The maximum contrast of the checkerboard pattern is on the Sr atoms,\cite{Marques2020} suggesting that it is linked to the second-nearest neighbour (NNN) hopping parameter in the tightbinding model, motivating a description through a staggered bond order (see Fig.~\ref{fig1}(c)). This NNN bond order breaks the translational symmetry on the Ru lattice, introducing a staggered next-nearest-neighbour interaction which is alternatingly strengthened and reduced by $\Delta_\mathrm{bond}$. The effect of the nematic and staggered bond order terms on the low energy electronic structure is comparable to the staggered on-site order proposed in Ref.~\onlinecite{Marques2020} and yields
four van Hove singularities in the density of states, see Fig.~\ref{Fig:Wannier}(d), very similar to the spectral features observed in tunneling, see Fig.~\ref{fig1} (b). The resulting electronic structure is shown in Suppl. Fig. 1/Suppl. Note 1. A description of the checkerboard charge order by the staggered on-site order yields similar results. We use the staggered bond order for the rest of the main text, but show key results for the staggered on-site order in Supplementary Note~3 and Supplementary Fig.~3. 

\subsection{Tunneling into $d_{xy}$ states}

In Fig. \ref{Fig:Wannier}(a), we plot cuts through the Wannier functions obtained via downfolding a DFT calculation of the unreconstructed surface of \ce{Sr2RuO4} onto a low-energy  band structure consisting of three $d$-orbitals. The full isosurfaces of these rather complicated functions are shown in Supplementary Fig.~2, but at the location of the STM tip, some $4-5\text{\AA}$ above the SrO surface plane, they resemble atomic $d$-orbitals. Note that the $d_{xz}$ Wannier function has one maximum roughly half way to the NN Ru atomic positions and the $d_{xy}$ Wannier function is much smaller in magnitude and vanishes by symmetry above the NN Ru atom. The octahedral rotation in the surface layer leads to a number of important changes in the electronic states associated with the $d_{xy}$ band: (1) the van Hove singularity in the $d_{xy}$ band shifts below the Fermi energy, (2) the $d_{xy}$ band acquires $d_{z^2}$ character with increased octahedral rotation and (3) the Wannier functions in the vacuum become chiral, with opposite chirality on Ru(1) and Ru(2) atoms. In Fig.~\ref{Fig:Wannier}(b), we show how the Wannier functions appear at the tip position in the presence of a 6$^\circ$ octahedral rotation. While the $d_{xz,yz}$ states are not qualitatively altered, the Wannier functions associated with the $\gamma$ band acquire a chiral character such that they no longer vanish above the NN Ru positions (light red dot).  The Wannier functions shown correspond to a Ru(2) position, with the function associated  with Ru(1) having the opposite chirality. Fig.~\ref{Fig:Wannier}(c) shows the electronic structure for an octahedral rotation of $\vartheta=6^\circ$ and the projected density of states for different octahedral rotations. The van-Hove singularity in the $d_{xy}$-derived $\gamma$ band has moved across the Fermi level compared to the unrotated case and has acquired a significant $d_{z^2}$ character, especially close to the M point, as a consequence of the octahedral rotation. The van-Hove singularity at the M point does not have any $d_{z^2}$ character in its projected density of states (PDOS) without rotation. These findings do not change in a fully relativistic ab-initio calculation.  It is through this admixture of $d_{z^2}$ character with the octahedral rotation that tunneling into the $\gamma$ band is facilitated and gives rise to peaks in the cLDOS from the vHss. These are absent when employing Wannier functions obtained without oxygen rotation (dashed lines in Fig. \ref{Fig:Wannier}(e)).

\begin{figure}
  \begin{center}
    \includegraphics[width=\linewidth]{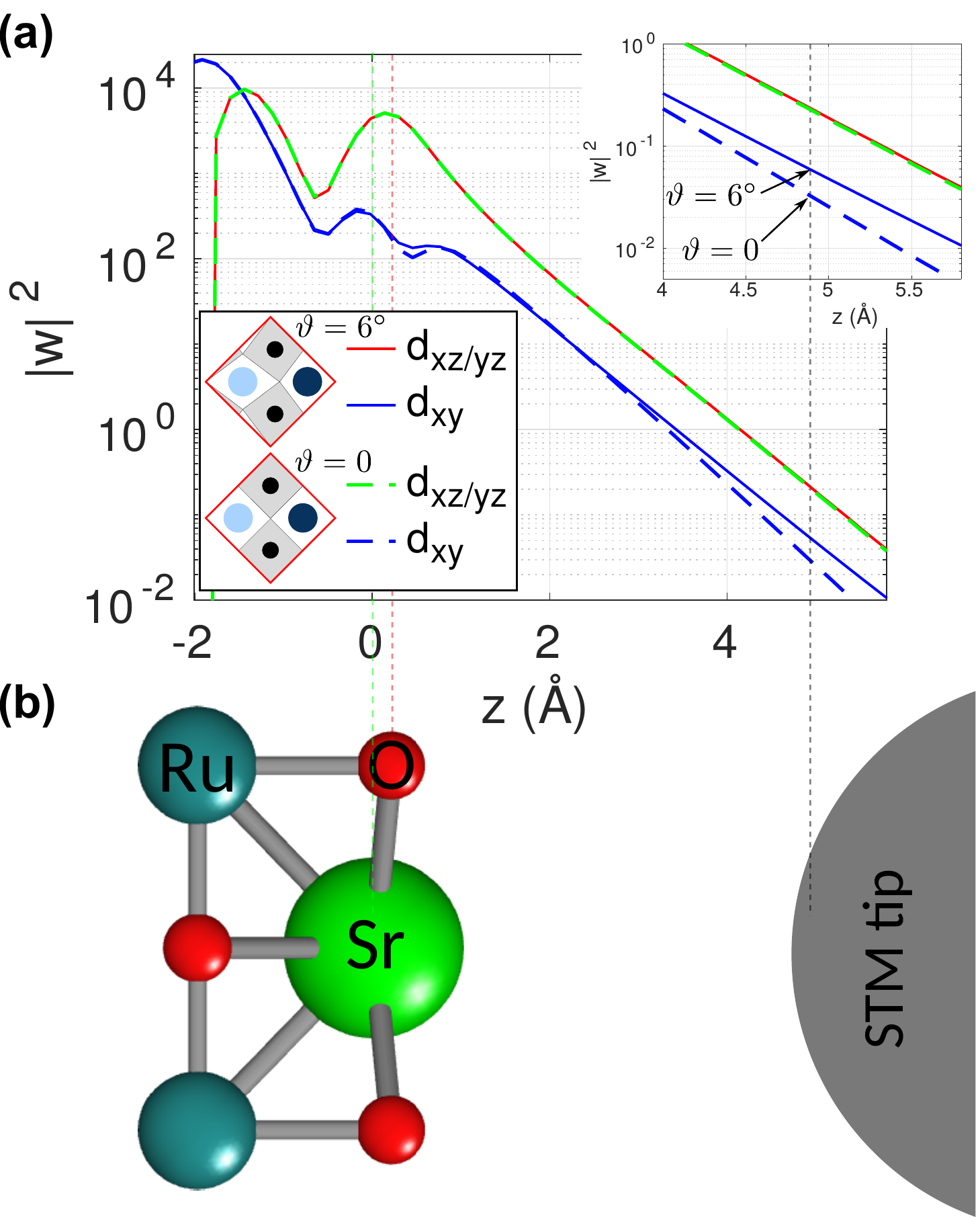}
    \end{center}
    \caption{\textbf{Tunneling probabilities at different tip-sample distances.} (a) Partial norm of the Wannier functions at fixed height $z$ showing the large values close to the atoms and the exponential decay in the vacuum. While the O octahedron rotation has negligible influence on the norm of the Wannier orbitals associated with the $d_{xz/yz}$ bands, it enhances the value of the Wannier orbital associated with the $\gamma$ band, which has predominantly $d_{xy}$ character, significantly, and also increases the decay length in $|W|^2\propto \exp(-z/\alpha)$ from $\alpha_{xy}=1.9\,\text{\AA}$ to $\alpha_{xy}=2.2\,\text{\AA}$. (b) Surface geometry of Sr$_2$RuO$_4$ with the Ru and O atoms on the surface at $z\approx 0$ (green and red dottet lines for these planes) and the STM tip approximately 5\AA{} above (black dashed line).
    }
    \label{Fig_wannier_decay}
\end{figure}
As a rough estimate of possible tunneling contributions, we plot in Fig.~\ref{Fig_wannier_decay}(a) the square of the Wannier function integrated over the $x-y$ plane, $|W|^2$, as a function of $z$, corresponding to the tip height in a tunneling experiment.
Once sufficiently away from the surface, for $z>2\text{\AA}$, the Wannier functions show the expected exponential decay, i.e. $|W|^2\propto \exp(-z/\alpha)$.
 
For the case without octahedral rotation, the $d_{xy}$ weight in vacuum at values of $z$ relevant for tunneling (assumed here to be typically at $5\text{\AA{}}$ above the surface, but the exact height is irrelevant for our analysis) is an order of magnitude smaller than the weight of $d_{xz,yz}$, as anticipated in Firmo et al.\cite{Firmo2013} However, once the octahedral rotation is considered, the weight of the $d_{xy}$ orbital is only about 3 times smaller and exhibits a decay length which is $10\%$ larger compared to the $d_{xz/yz}$ orbitals (decay length $\alpha_{xy}=2.2 \text{\AA}$ vs. $\alpha_{xz,yz}=2.0 \text{\AA{}}$). The decay length for the $d_{xy}$ orbital thus changes from a value smaller than that of the $d_{xz,yz}$ orbitals to a larger value due to the rotation. 
The suppression of the vacuum overlap in the unreconstructed state alone is therefore not sufficient to explain the lack of most $d_{xy}$ features in QPI. We will show below that the $d_{xy}$ states contribute most strongly near the van Hove point, close to ${\bf k}=(\frac{1}{2},0)$, thus tunneling should be proportional to the value of the Wannier functions at the centre of the NN Ru atom. Without rotation of the oxygen octahedra, this is zero by symmetry, see Fig.~\ref{Fig:Wannier}(a), i.e. no tunneling is expected from states close to ${\bf k}=(\frac{1}{2},0)$. This is also seen by the absence of any features due to the vHs in the cLDOS when calculated without rotation of the oxygen octahedra, see Fig.~\ref{Fig:Wannier}(e).

\begin{figure}
    \begin{center}
    \includegraphics[width=\linewidth]{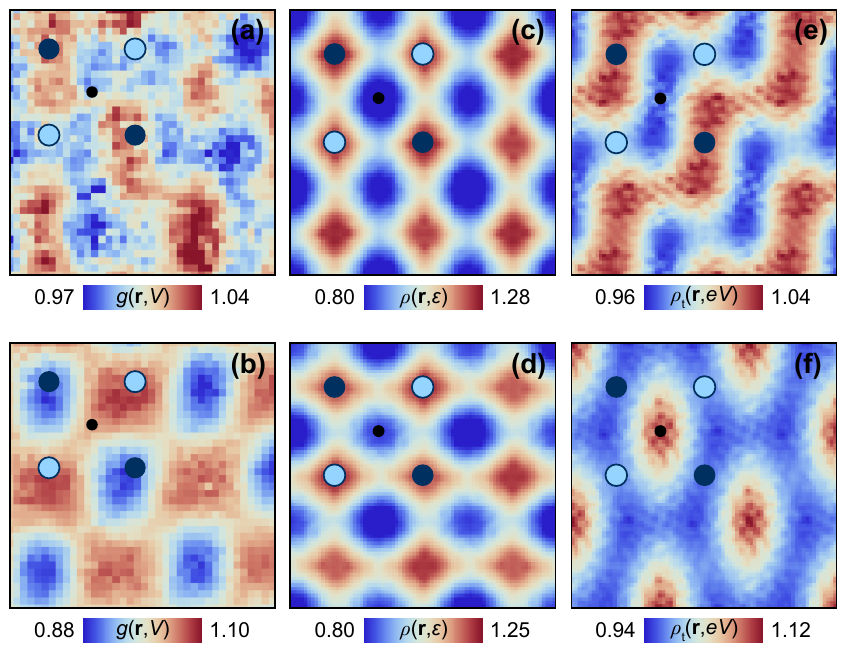}
    \end{center}
\caption{\textbf{Checkerboard charge order.} (a, b) Experimental differential conductance maps $g(\mathbf{r},V)$ taken at the energy of the vHs\cite{Marques2020} at (a) positive energy, $V=3.5\mathrm{mV}$ and (b) negative energy, $V=-3.5\mathrm{mV}$ at which the checkerboard order is most prominent ($T=59\mathrm{mK}$, $V_\mathrm{set}=7.0\mathrm{mV}$, $I_\mathrm{set}=250\mathrm{pA}$, $V_\mathrm{L}=495 \mathrm{\mu V}$). (c, d) cLDOS $\rho(\mathbf{r}, \epsilon)$ calculated at a height $z=5\text{\AA}$ above the surface at (c) $\epsilon=5\mathrm{meV}$ and (d) $\epsilon=-5\mathrm{meV}$. (e, f) calculated differential conductance map $\rho_\mathrm{t}(\mathbf{r},eV)$ at (e) $V=5\mathrm{mV}$ and (f) $V=-5\mathrm{mV}$, emulating the effect of the feedback loop of the STM (see main text) for $E_\mathrm{set}=10\mathrm{meV}$. 
The intensity of all images has been normalized by the spatial average, color bars indicate relative intensity. Filled dark and light blue circles mark the positions of Sr atoms, black filled circles of the Ru atoms, as in Fig.~\ref{Fig:Wannier}(c). For experimental details, see Appendix \ref{app_stm}.}
\label{fig:checkerboard}
\end{figure}
 
\subsection{Checkerboard charge order}
In experiments, one of the most prominent features of the surface electronic structure is a pronounced Sr-centred checkerboard charge order, which in our model is accounted for through the staggered bond order. In Fig.~\ref{fig:checkerboard}(a) and (b) we show measured  differential conductance maps at positive and negative bias voltages, respectively, in comparison to calculated maps (Fig.~\ref{fig:checkerboard}(c, d)) of the continuum local density of states at a constant height above the surface, fully accounting for the vacuum tail of the wave functions. The LDOS maps demonstrate that the staggered bond order indeed leads to a checkerboard charge order centred on the Sr atoms as found experimentally, and reproduces the contrast inversion between positive and negative bias voltages (compare Fig.~\ref{fig:checkerboard}(c, d)). A notable difference between the experimental and calculated maps is that the experimental data is dominated by the checkerboard charge order, whereas the calculated maps show the checkerboard charge order as a subdominant contribution superimposed to the atomic contrast. We attribute this difference between the experimental and calculated maps to the different treatment of the tip-sample distance: in our measurements, the tip-sample distance is set at each point independently to yield a constant current, whereas in the calculations shown in Fig.~\ref{fig:checkerboard}(c, d) the local density of states is taken at constant height. To faithfully reproduce the experimental data requires calculating differential conductance maps where the tip height is locally adjusted to maintain a constant integral $\int_0^{E_\mathrm{set}}\rho(E)\mathrm dE$, emulating the effect of the feedback loop of an STM regulating on a constant tunneling current before the spectrum is recorded. Such maps are shown in Fig.~\ref{fig:checkerboard}(e, f) for the same energies as in (c, d), showing a complete suppression of the atomic contrast. At positive bias voltages, we find excellent agreement between the calculated and measured differential conductance maps (Fig.~\ref{fig:checkerboard}(a, c)), whereas at negative bias voltage the agreement is not quite as good: The LDOS map in Fig.~\ref{fig:checkerboard}(d) reproduces the checkerboard charge order as seen experimentally, but the calculated differential conductance map (Fig.~\ref{fig:checkerboard}(f)) shows the dominant contrast on top of the ruthenium atoms. This difference is likely an artifact because the simulated tip-sample distance is significantly smaller than the one expected for the experiment.
The theoretically tractable tip-sample distances are limited due to technical reasons related to the accuracy of the Wannier functions at large distances, indications for quantitative changes are given by the (slightly) larger decay length of the $d_{xy}$ Wannier function.
Our results show that the setpoint effect, which is normally considered detrimental to the interpretation of spectroscopic maps, suppresses the atomic corrugation in the differential conductance maps. For the following comparison of the quasi-particle interference, we have verified that the main impact of the setpoint effect is a suppression of the atomic contrast in the differential conductance maps, otherwise not affecting the signal due to quasi-particle interference significantly. For comparison, we show in the supplementary material (Suppl. Fig. 3 and Suppl. Note 3) the calculations for the staggered onsite order, showing very similar results.

\begin{figure}[t]
    \begin{center}
    \includegraphics[width=\linewidth]{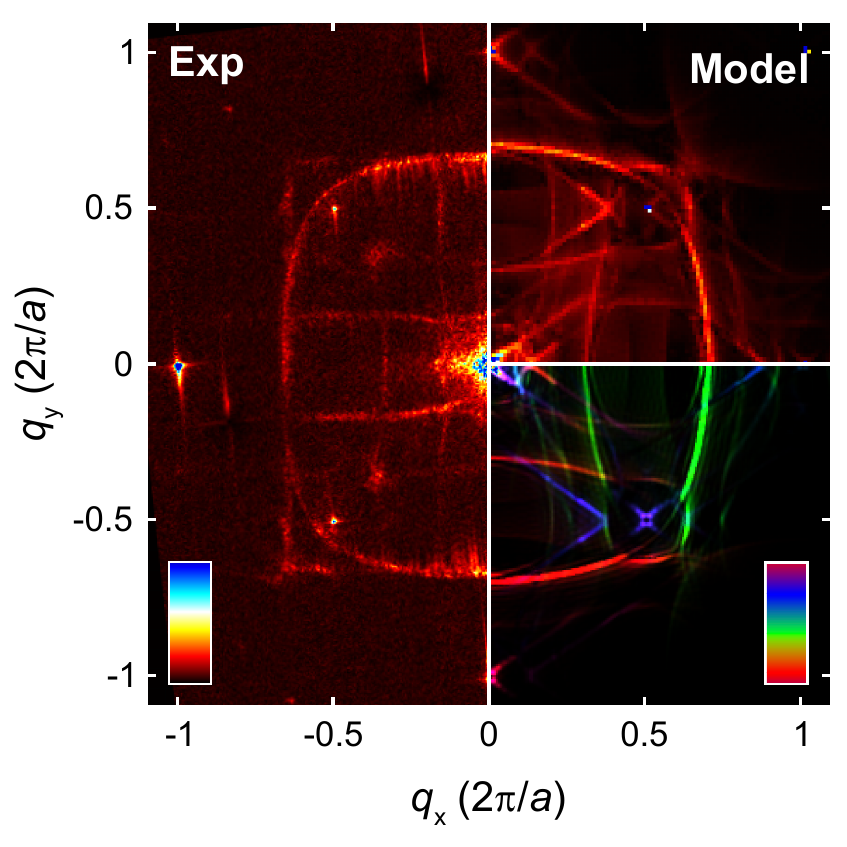}
    \end{center}
    \caption{\textbf{Quasiparticle Interference.} Comparison of an experimental QPI map on the left ($V_\mathrm{set}=5.8\mathrm{mV}$, $I_\mathrm{set}=200\mathrm{pA}$, $T=600\mathrm{mK}$, $B=12\mathrm{T}$, aliased Fourier peaks have been suppressed for clarity) with the theoretically calculated QPI map on the top right obtained from Fourier transformation of the real-space continuum LDOS at $2\mathrm{meV}$, averaging over scattering patterns from both types of Ru impurities (see Suppl. Fig. 4 for real space patterns close to individual Ru-site defects, $V_\mathrm{imp}=+0.1\mathrm{eV}$, $z=5\text{\AA{}}$). The bottom right panel shows the orbital decomposition for the cLDOS calculations with red representing $d_{yz}$-, green $d_{xz}$- and blue $d_{xy}$-character.}
    \label{fig3}
\end{figure}

\subsection{Quasiparticle Interference}
In order to provide a full picture of the low energy electronic structure around the Fermi energy, we use QPI imaging and compare our experimental QPI maps to the simulated continuum LDOS maps. Continuum LDOS maps in real space and simulated topographies exhibit chiral QPI patterns around Ru-site defects, as also found experimentally. The rotational sense of these patterns depends on the position of the defect in a Ru(1) or Ru(2) site (see Supplementary Fig.~4 and Supplementary Note 4). For comparison with the experiment, we average over both types of defects, as also the experimental data is acquired over fields of view with defects in both sites. Comparison of the simulated QPI, fully taking into account the tunneling matrix elements through the Wannier functions (for details on the method see Appendix \ref{app_wannier}),
reveals excellent overall agreement between experiment and theory (see Fig.~\ref{fig3}).
We note that the positions of features in $\mathbf{q}$-space deviate slightly between theory and experiment. For example the outer dominant ring-like structure is larger in the calculation. This is because the bare electronic structure from the first principles calculations does not to match exactly the true Fermi surface, a deviation which is not relevant for the following discussion. The key features are qualitatively consistent between theory and experiment: the outer square-shaped scattering from the quasi-1D bands, and the inner square coming from the scattering processes crossing the zone-boundary. At low $\mathbf{q}$-vectors, an intensity distribution with $C_2$ symmetry is seen which is reproduced in the calculation.
As pointed out previously\cite{Wang2017}, the appearance of spectral weight in some parts of the BZ can only be understood accounting for the surface reconstruction; these are essentially the structures parallel to $\mathbf{q}_x$ and $\mathbf{q}_y$ in Fig.~\ref{fig3}, which can be traced back to bands with predominantly $d_{xz}$/$d_{yz}$ character (see also orbital decomposition in fig.~\ref{fig3}). By contrast, as can be seen in Fig.~\ref{fig3} there are scattering processes close to the atomic and reconstruction peaks associated with the $d_{xy}$-derived $\gamma$-band, the band which exhibits the van Hove singularity close to the Fermi energy.
In the following, we will use the model calculations to establish the signatures of the $\gamma$-band and the van-Hove singularity in QPI.

\begin{figure*}[th!]
    \begin{center}
    \includegraphics[width=\linewidth]{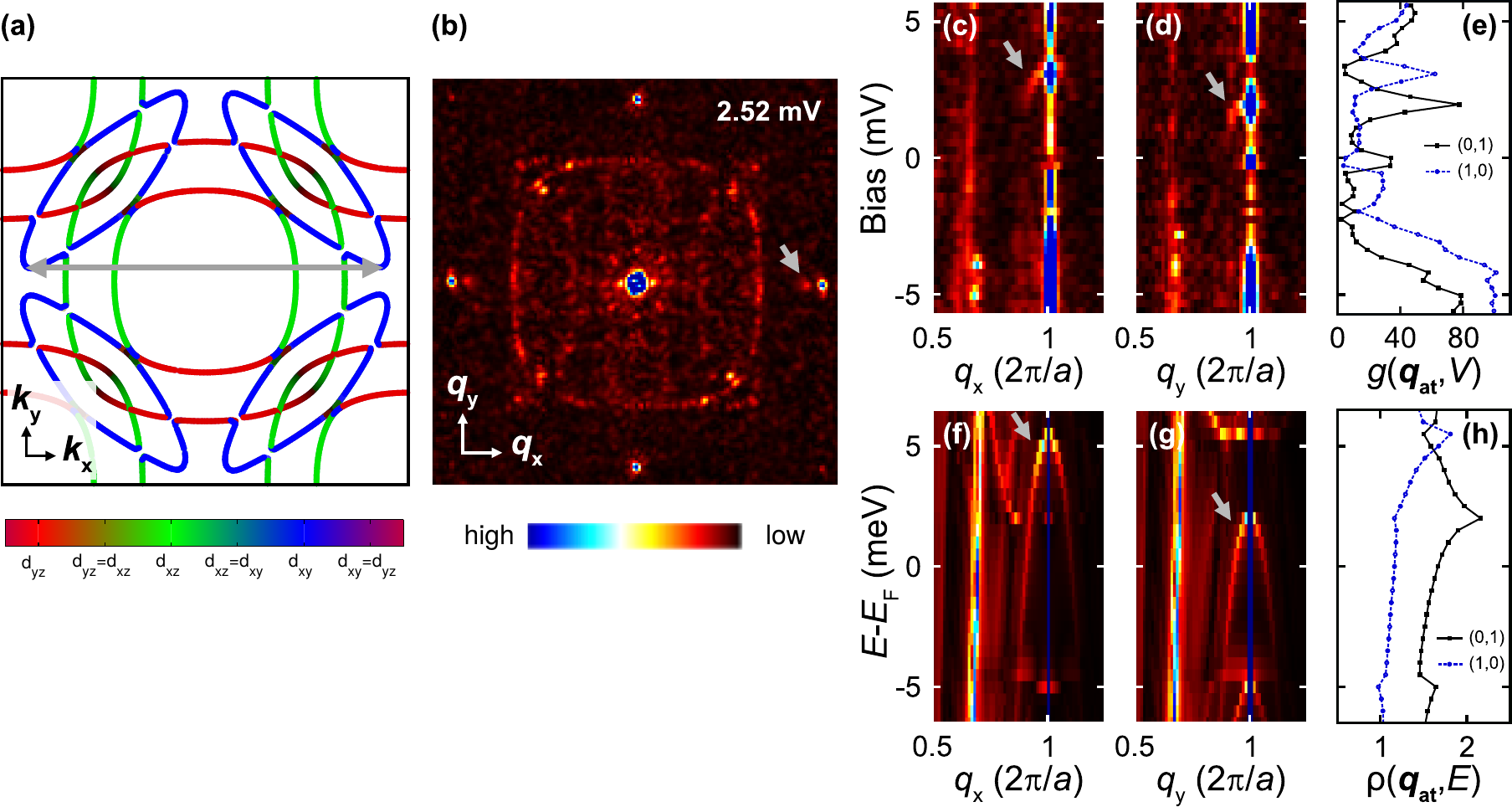}
    \end{center}
    \caption{\textbf{QPI close to the atomic peaks.} (a) Fermi surface from the model with surface reconstruction. The color represents the orbital character of the bands, with $d_{xz}$/$d_{yz}$ character shown in green/red, respectively and $d_{xy}$ character in blue. The grey arrow indicates the $\mathbf{q}$-vector connecting the tips of the pockets ofthe $\gamma$-band close to the van Hove singularity. The QPI dispersion from this $\mathbf{q}$-vector is expected near the atomic peaks. (b) Fourier transformation of a differential conductance map at $V=2.52 \mathrm{mV}$. QPI features due to the $\mathbf{q}$-vector shown in (a) (marked by an arrow) can be observed close to the Bragg peaks at $(1,0)$ (in units of $2\pi/a$). (c, d) Energy-momentum cuts through the QPI map along $\mathbf{q}_x$ (c) and $\mathbf{q}_y$ (d) close to $(1,0)$ and $(0,1)$. A clear dispersing feature is seen which collapses onto the atomic peak (white arrows). The dispersion differs in the $\mathbf{q}_x$ and $\mathbf{q}_y$ directions as a consequence of nematicity ($V_\mathrm{set}=5.6$ mV, $I_\mathrm{set}=225$ pA, $V_\mathrm{L}=300$ $\mu$V, $T=76$ mK, $B=6.5$ T). (e) Differential conductance $\tilde{g}(\mathbf{q}_\mathrm{at},V)$ for the atomic peaks $\mathbf{q}_\mathrm{at}=(1,0)$ and $(0,1)$, showing prominent features at $V=2\mathrm{mV}$ and $3\mathrm{mV}$ where the vHs crosses the zone boundary. (f, g) Corresponding energy-momentum cuts from the model along $\mathbf{q}_x$ (f) and $\mathbf{q}_y$ (g), showing the signatures of the dispersion of the scattering vectors associated with the $\gamma$ band around the atomic peaks. (h) As in (e), Local density of states $\tilde{\rho}(\mathbf{q}_\mathrm{at},E)$ at $\mathbf{q}_\mathrm{at}=(1,0)$ and $(0,1)$ as a function of $E$. As in the experimental data, two maxima are seen due to the vHs at the zone boundary.}
    \label{fig4}
\end{figure*}

\subsubsection{QPI of the van Hove singularity}
The QPI signal of the $\gamma$-band and the vHs is dominated by scattering vectors connecting the tips of the constant energy contours close to the vHs. Due to the background near $q\rightarrow 0$ from the impurity distribution, the small $\mathbf{q}$ vectors connecting the tips near the vHs are difficult to detect reliably. We have identified two ways to still accurately detect the dispersion close to the vHs and determine the energy of the vHs: (1) the scattering vectors connecting the points of highest density of states include ones which cross the Brillouin zone, leading to QPI features around the atomic Bragg peaks, where the noise background is much lower. (2) At the van Hove singularity, the scattering vector becomes commensurate with the atomic contrast, resulting in a resonant enhancement of the atomic contrast when the energy becomes equal to that of the vHs.\\
Fig.~\ref{fig4}(a) shows the Fermi surface extracted from our tight-binding model, with one of the scattering vectors with high joint density of states connecting points near the van-Hove singularity leading to QPI around the atomic peaks. This scattering vector is already apparent from the QPI map as a feature in close proximity to the atomic peaks, see Fig.~\ref{fig4}(b). From line cuts through the three-dimensional energy-momentum data along the $\mathbf{q}_x$ and $\mathbf{q}_y$ direction, Fig.~\ref{fig4}(c, d), the dispersion of these peaks can be tracked. A clear hole-like dispersion is observed with a band maximum a few millivolts above the Fermi energy. Fig.~\ref{fig4}(f, g) show for comparison the same cuts obtained from the calculations. While the calculations exhibit more fine structure than seen in the experiment, the main feature of a hole-like dispersion around the atomic peak is faithfully reproduced. As expected from the nematicity of the electronic structure, the van-Hove singularities occur at slightly different energies along the $\mathbf{q}_x$- and $\mathbf{q}_y$- directions, providing an estimate of the magnitude of the nematic term $\Delta_\mathrm{nem}$ in the Hamiltonian. 
The nematicity also leads to a pronounced anisotropy of the low-$q$ QPI. As a function of energy, the contributions from small $q$ scattering vectors crossing the zone boundary are expected to evolve according to the touching of the (reconstructed) bands. In real space, these correspond to interference patterns with long wavelength and rotation of the dominant wave vector as a function of energy, as noted in Ref. \onlinecite{Marques2020}. Calculated maps of the continuum LDOS confirm this feature and allow us to unequivocally assign it to tunneling into the $d_{xy}$-derived $\gamma$-band  since the van Hove singularity responsible for this low-$q$ scattering occurs only in this orbital channel.
Notably, these results confirm our theoretical conjecture that the $\gamma$ band becomes detectable in tunneling, facilitated by the octahedral rotations.
 
The crossing of the QPI signal of the dispersion of the $\gamma$-band through the atomic peak seen in Fig.~\ref{fig4}(c, d) provides an alternative measure of the van Hove singularities in the electronic structure: at the energy of a vHs at the zone boundary, the quasi-particle scattering becomes commensurate with the atomic periodicity leading to a significant increase in the intensity of the atomic peaks in spectroscopic maps. In Fig.~\ref{fig4}(c, d), this becomes apparent as saturation of the contrast at distinct points along the $\mathbf{q}_\mathrm{at}=(1,0)$ and $(0,1)$ line. For clarity, we plot the energy dependence of the QPI signal $\tilde{g}(\mathbf{q}_\mathrm{at},V)$ as a function of energy $eV$ in Fig.~\ref{fig4}(e). Traditional (lattice-only) T-matrix calculations are unable to capture this feature because they exhibit the same periodicity as the Brillouin zone, and the QPI signal at the atomic peak is thus identical to the one at the zone centre. The continuum LDOS is able to describe this intensity modulation: in our calculations, the intensities of the atomic peaks at $\mathbf{q}_\mathrm{at}$ in maps of the LDOS $\tilde{\rho}(\mathbf{q},E)$ show sharp peaks when the $\gamma$ band crosses the zone boundary (compare Fig.~\ref{fig4}(h)), providing an alternative way to determine the energy of the van Hove singularities in the electronic structure without the necessity of undertaking a full QPI mapping. 

\section{Discussion}
Our theoretical modelling and measurements provide a comprehensive picture of the low energy electronic structure of the surface layer of \ce{Sr2RuO4}, and identify clear signature of the $\gamma$-band in QPI with potential implications for its superconducting state. Quasi-particle interference of the $\gamma$-band which is predominantly of $d_{xy}$-character had hitherto been assumed to only contribute negligibly to the tunneling signal. Our measurements show a clear QPI signal from this band through comparison with theory.
From the calculations for a hypothetical unreconstructed surface, without octahedral rotation, we indeed find that tunneling to the $\gamma$ band would be about an order of magnitude smaller than for the $d_{xz}$ and $d_{yz}$ bands, leading to a negligible contribution to the tunneling conductance. The small tunneling probability is due to the real space properties of the $d_{xy}$ Wannier function and the oscillatory nature of the Bloch wave function near $\mathbf{k}=(\frac{1}{2},0)$. 
In the reconstructed surface, the octahedral rotation leads to additional $d_{z^2}$ weight for the $\gamma$-band, making it accessible by QPI.

In summary we have identified the physical ingredients necessary to describe the electronic structure at the surface of \ce{Sr2RuO4} which include spin-orbit coupling, the nematic and staggered charge orders as well as the rotation of the oxygen octathedra. While the terms in the Hamiltonian outlined in detail in the Methods section can in principle be derived from a microscopic description, the effective parameters are subject to renormalizations due to the strongly correlated nature of the material.
Despite the excellent agreement we observe between the experimental data and theoretical modelling, one notable difference remains: while in tunneling spectroscopy the gap-like structure around the Fermi energy leads to a significant suppression of differential conductance by about $40\%$, this is not accurately captured in the calculation, where the suppression remains significantly smaller. This can have a number of origins, including that the calculations are carried out for smaller tip-sample distances than used in the experiments, potential additional relaxation of the surface layer in the $z$-direction and that a larger part of the Fermi surface becomes gapped out than is captured in the model. Nevertheless, our measurements clearly demonstrate that all three bands of the $t_{2g}$ manifold ($d_{xz}$, $d_{yz}$ and $d_{xy}$) which are expected to be present at the Fermi energy can be detected in QPI. We also note that our measurements, taken at 59 mK, do not show evidence of a superconducting gap. This is particularly surprising given that all three bands which contribute to the Fermi surface are clearly detected. The absence of spectral features from such a gap in many high-resolution STM experiments on the SrO-terminated surface of \ce{Sr2RuO4}\cite{kambara_scanning_2006, lupien_mk-stm_2011,  Marques2020} remains an important open puzzle. One possibility highlighted by our analysis is that superconductivity is suppressed if the surface is reconstructed by octahedral rotation, as assumed here.  It is conceivable that disorder, or other subtle surface effects, may lead to other reconstructions that do not suppress superconductivity -- calling for new ways to suppress the surface reconstruction, possibly through adsorbate layers to facilitate a detection of the superconducting gap in tunneling experiments and thus determination of the superconducting order parameter -- providing a resolution to the long-standing mystery of the symmetry of the superconducting order parameter in \ce{Sr2RuO4}.\\

\appendix
\section{Methods}
\subsection{First principles Wannier calculations}
\label{app_dft}
Density functional theory calculations~\cite{hohenberg1964inhomogeneous} were performed with the projected augmented wave (PAW) method as implemented in the Vienna \textit{ab initio} simulation package (VASP)~\cite{PhysRevB.54.11169,PhysRevB.59.1758}. The generalized gradient approximation of Perdew, Burke and Ernzerhof was used for the  exchange correlation functional~\cite{PhysRevLett.77.3865}. To be able to capture the Wannier functions high above the surface we performed a monolayer calculation of perovskite Sr$_2$RuO$_4$. The lattice constant was taken $a=3.87\text{\AA}$, and the vacuum length was chosen to be about $21\text{\AA}$. To incorporate the rotations we construct a $\sqrt{2}\times\sqrt{2}$ supercell of the Sr$_2$RuO$_4$ monolayer unit cell. We perform two calculations, one without a rotation and one with a $\vartheta=6^{\circ}$ rotation of the Ru-O in-plane bonds.
In the rotation of the O atoms we keep the lattice constants, the Ru positions and the Ru-O bond angles fixed suth that the O position is then given by $(x,y,0)=0.25(\tan\vartheta+1,\tan\vartheta-1,0)$. The energy cutoff of the plane waves was chosen as 650 eV. The total energy was converged to 10$^{-7}$ eV. The Brillouin zone integration was sampled by using a $7\times7\times1$ $\Gamma$-centered Monkhorst-Pack grid. To construct the Wannier functions and the tight-binding models, the $d_{xz}$, $d_{yz}$ and $d_{xy}$ orbitals were projected on the low energy bands, employing the Wannier90 code package~\cite{mostofi2014updated} with input parameters \verb+num_iter=0+ and \verb+dis_num_iter=10000+. The outer energy window was taken as [-3,1] eV and the inner frozen energy window as [-1.7,0.2] eV, both relative to the Fermi level. 
\subsection{Surface tight-binding Hamiltonian with spin-orbit coupling}
\label{app_tb}
As described in the main text, the Hamiltonian matrix in momentum space is given by
\begin{equation}
 H(\mathbf{k})=H_{\mathrm{tb}}(\mathbf{k})+H_{\mathrm{soc}}
\end{equation}
where
$H_{\mathrm{tb}}(\mathbf{k})=1_{\mathrm{spin}}\otimes t(\mathbf{k})^{ab}$ is the Bloch Hamiltonian as obtained from the ab initio Wannier calculation (including the $\vartheta=6^{\circ}$ rotation of the Ru-O in-plane bonds) with an overall band renormalization of $Z=1/4$ to match experimentally observed Fermi velocities\cite{Tamai2019}. The $6\times 6$ matrix $t(\mathbf{k})^{ab}$ is the Fourier representation of the in-plane hoppings of our two dimensional model for the orbitals $a$ and $b$ with $a,b=(d^{(1)}_{xz},d^{(1)}_{yz},d^{(1)}_{xy},d^{(2)}_{xz},d^{(2)}_{yz},d^{(2)}_{xy})$ and the superscript denotes the sublattice index of the Ru atoms.

We introduce spin-orbit coupling via an onsite spin-orbit coupling term proportional to the unit matrix in the sublattice space,
\begin{equation}
 H_{\mathrm{soc}}=\lambda \left(\begin{array}{cccc}
  H^{(1)}_{\up\up}& 0&H^{(1)}_{\up\down}&0\\
  0& H^{(2)}_{\up\up}& 0&H^{(2)}_{\up\down}\\
  H^{(1)}_{\down\up}&0&H^{(1)}_{\down\down}&0\\
  0 & H^{(2)}_{\down\up}& 0&H^{(2)}_{\down\down}
  \end{array}\right)
\end{equation}
where the $3\times 3$ matrices stem from the product of the spin operator $\mathbf S=\frac 1 2 (\sigma_x,\sigma_y,\sigma_z)$  and the representation of the angular momentum matrices $L_\mu$ in the basis of the real-valued $d$ orbitals, $H^{(s)}_{\up\up}= L_z/2$, $H^{(s)}_{\up\down}=(L_x+iL_y)/2$, $H^{(s)}_{\down\up}=(L_x-iL_y)/2$, $H^{(s)}_{\down\down}=- L_z/2$ for $s=1,2$, where
\begin{align}
 L_x=\left(\begin{array}{ccc}
            0& 0&i\\
            0&0&0\\
            -i &0&0
           \end{array}\right)\\
 L_y=\left(\begin{array}{ccc}
            0&0&0\\
            0&0&i\\
            0&-i&0
           \end{array}\right)\\
  L_z=\left(\begin{array}{ccc}
            0&-i&0\\
            i&0&0\\
            0&0&0
           \end{array}\right)\,.
\end{align}
We use $\lambda=20\,\mathrm{meV}$ to yield agreement of the splittings from the avoided crossings of the quasi 1 dimensional $d_{xz}$ and $d_{yz}$ bands along the path from $\Gamma$ to the M point (see Fig. \ref{fig4}) with the measured spectral function of the surface bands\cite{Tamai2019}.
\subsection{Nematicity and Staggered orders}
\label{app_nem}
We introduce the nematic and staggered orders through an additional term $H_{\mathrm{nem}}(\mathbf{k})$ in the Hamiltonian,
\begin{equation}
 H(\mathbf{k})=H_{\mathrm{tb}}(\mathbf{k})+H_{\mathrm{soc}}+H_{\mathrm{nem}}(\mathbf{k}).
\end{equation}
The nematic and bond order contributions are diagonal in spin space, $H_{\mathrm{nem}}(\mathbf{k})=1_{\mathrm{spin}}\otimes t_{\mathrm{nem}}(\mathbf{k})$. The matrix $t_{\mathrm{nem}}(\mathbf{k})$ has only nonzero components in the  $d_{xy}$ orbital components such that the sub-matrix in this subspace, spanned by the elements 3 and 6, reads
\begin{equation}
 t_{\mathrm{nem}}(\mathbf{k})|_{[3,6],[3,6]}=\left(\begin{array}{cc}
 \Delta_\mathrm{bond} f_\mathrm{bond}(\mathbf{k}) &  \Delta_\mathrm{nem}f_\mathrm{nem}(\mathbf{k})\\
 \Delta_\mathrm{nem}f_\mathrm{nem}^*(\mathbf{k}) &  -\Delta_\mathrm{bond} f_\mathrm{bond}(\mathbf{k})
                                                  \end{array}\right)
\end{equation}
with $f_\mathrm{bond}(\mathbf{k})=\frac 1 2 (\cos k_x+\cos k_y)$ and $f_\mathrm{nem}(\mathbf{k})=\frac 1 4[1+\exp[i(k_x-k_y)]-\exp(-i k_y)-\exp(ik_x)]$
where the momenta are in the Brillouin zone obtained from an elementary cell with two Ru atoms, see Fig.~\ref{fig1}(c) and we chose  $\Delta_{\mathrm{nem}}=2.5\,\mathrm{meV}$, $\Delta_\mathrm{bond}=5\,\mathrm{meV}$. To describe the staggered on-site order, we use the same expression with $f_\mathrm{bond}(\mathbf{k})=1$.
\subsection{Wannier method for simulations of tunneling}
\label{app_wannier}
Here we adopt a Wannier-function based approach which allows us to relate the tunneling rate to the local density of states in vacuum a few \AA{} above the surface of the SrO layer, where the STM tip is positioned.
The current at voltage $V$ (or differential conductance) as measured in an STM experiment  can be calculated by
\begin{equation}
 I(V,\mathbf{r})=
 A_0
 \int_0^{eV} \rho(\mathbf{r},\omega) d\omega \,
 \label{eq_tunneling}
\end{equation}
where  $A_0$ is a constant containing the tip density of states and the tunneling matrix element, and $\rho(\mathbf {r},\omega)$ is the continuum local density of states (cLDOS) at the tip position $\mathbf{r}=(x,y,z)$ which is assumed to be at several \AA{} above the surface atoms of \ce{Sr2RuO4}.
The cLDOS can be calculated conveniently from the continuum Green's function $G(\mathbf {r},\mathbf {r}';\omega)=\langle \psi(\mathbf{r})^\dagger \psi(\mathbf{r}')\rangle_\omega$ (where $\psi(\mathbf{r})=\sum_{\mathbf{R},\mu}c_{\mathbf{R}\mu}w_{\mathbf{R}\mu}$ are the continuum electron operators) via \cite{Choubey14,Chi2016A}
\begin{equation}
 \rho(\mathbf {r},\epsilon)=-\frac 1 \pi \mathop{\text{Im}}G(\mathbf {r},\mathbf {r};\epsilon).\label{eq_cldos}
\end{equation}
Usually, the electronic structure is discussed using the lattice (tight-binding) model with a lattice Green's function ${\hat G}_{\mathbf{R}, \mathbf{R'}}^{\mu,\nu}(\omega)$, a matrix in the combined orbital and spin space $\mu$, lattice position $\mathbf{R}$.
The continuum Green function can be calculated using a basis transformation as
\begin{equation}
 {G}(\mathbf{r},\mathbf{r}';\omega)=\sum_{\mathbf{R}, \mathbf{R'},\mu\nu}{\hat G}_{\mathbf{R}, \mathbf{R'}}^{\mu,\nu}(\omega)w_{\mathbf{R}\mu}(\mathbf{r})w_{\mathbf{R'}\nu}(\mathbf{r'}),
 \label{eq_basis}
\end{equation}
where the matrix elements $w_{\mathbf{R}\mu}(\mathbf{r})$ are the Wannier functions which are obtained in the tight-binding downfolding for $H_{\mathrm{tb}}(\mathbf{k})$ as well. Finally, let us mention that the differential
conductance (at constant tip height) is obtained by taking the derivative of  Eq. (\ref{eq_tunneling}) with respect to the bias voltage, yielding the proportionality $dI/dV\propto \rho(\mathbf{r},eV)$.\renewcommand{\underline}{}
A topographic map $z(x,y)$ as obtained experimentally by keeping the current $I_0$ constant for a given bias voltage $V_0$, can be calculated by solving the equation
\begin{equation}
I_0=A_0\int_0^{eV_0} d\omega~  \rho({x,y,z(x,y)},\omega)\,,\label{eq_topograph}
\end{equation}
for $z(x,y)$ which requires the evaluation of the continuum LDOS within a height range and for all energies to carry out the integral.
Finally, for realistic calculations of conductance maps as obtained in topographic mode, one evaluates continuum LDOS at the height profile\cite{Choubey17}, i.e,
\begin{equation}
\rho_{\mathrm{t}}(x,y,eV)=\rho((x,y,z(x,y)),eV)\,.
\end{equation}
In a homogeneous system, the lattice Green function is translation invariant, ${\hat G}_{\mathbf{R}, \mathbf{R'}}=\underline{\hat G}_{\mathbf{R}-\mathbf{R'}}^0(\omega)$, where the r.h.s. can be calculated by Fourier transform from the Green function in momentum space \mbox{$\underline{\hat G}_{\mathbf{k}}^0(\omega)=[H(\mathbf{k})-\omega]^{-1}$}, $\underline{\hat G}_{\mathbf{R}}^0(\omega)=\sum_{\mathbf{k}} \underline{\hat G}_{\mathbf{k}}^0(\omega) e^{i\mathbf{R}\cdot\mathbf{k}}$.

For calculations including impurities, we consider a simple potential scatterer at a Ru atom with lattice position $i^*$, diagonal in the combined orbital (sublattice) and spin space with the Hamiltonian
\begin{equation}
  H_{\mathrm{imp}}= V_{\mathrm{imp}}\sum_{\alpha} c_{i^*,\alpha}^\dagger c_{i^*,\alpha}.
  \label{eq_H_imp}
\end{equation}
Within the T-matrix approach, the lattice Green function is given by\cite{Chi2016A,Choubey17}
\begin{equation}
 \underline{\hat G}_{\mathbf{R}, \mathbf{R'}}(\omega)=\underline{\hat G}_{\mathbf{R}-\mathbf{R'}}^0(\omega)+\underline{\hat G}_{\mathbf{R}}^0(\omega)\underline{\hat T}(\omega)\underline{\hat G}_{-\mathbf{R'}}^0(\omega)\,,
 \label{eq_lattice_GF_T}
\end{equation}
where 
\begin{equation}
 \underline{\hat T}(\omega)=[1-\underline{\hat V}_{\text{imp}}\underline{\hat G}(\omega)]^{-1} \underline{\hat V}_{\text{imp}},
 \label{eq_T_matrix}
\end{equation}
is the T-matrix and $\underline{\hat V}_{\text{imp}}= V_{\mathrm{imp}} 1_{\mathrm{spin}}\otimes \hat S$ is the matrix representation of Eq.(\ref{eq_H_imp}). For an impurity on a Ru(1) atomic position, we use {\small
$ \hat S=\left(\begin{array}{cc}
               1_3&0\\
               0&0\end{array}\right)$}, while for the impurity on a Ru(2) atomic position {\small
$ \hat S=\left(\begin{array}{cc}
               0&0\\
               0&1_3\end{array}\right)$}.
The local Green function is just given by $\underline{\hat G}(\omega)=\underline{\hat G}_{\mathbf{R}=0}^0(\omega)$. Note that all frequency arguments in the Green functions are shorthand notations for $\omega+i\eta$ with an energy broadening $\eta$ which we choose to be sub meV to achieve satisfactory energy resolution by use of k grids of size $2500\times 2500$ (or $3500\times 3500$ for Fig. \ref{Fig:Wannier}(d,e)) . Finally, lattice density of states can be calculated as trace, $N({\bf q},\omega)=-\frac 1 \pi \mathop{\mathrm{Im}} \mathop{\mathrm{Tr}}\underline{\hat G}_{\mathbf{q}}^0(\omega)$.\\

\subsection{Ultra-low temperature STM}
\label{app_stm}
Quasi-particle interference imaging has been performed using a dilution-refrigerator based low temperature STM operating at temperatures down to $10\mathrm{mK}$\cite{Singh2013}. Spectroscopic maps $g(\mathbf{r},V)$ shown here, where a tunneling spectrum is recorded at each point of a topographic image $z(\mathbf{r})$, were recorded by stabilizing the tip sample distance at a tunneling setpoint $V_\mathrm{set}$, $I_\mathrm{set}$ before switching the feedback loop off to record the spectrum. The differential conductance was recorded through a lock-in technique, adding a voltage modulation $V_\mathrm{L}$ to the bias voltage. While some of the maps shown here where recorded in magnetic field (as indicated in figure captions), the qualitative behaviour of the features reported here is not affected by the field. The magnetic field shifts some features slightly in energy (as reported in \cite{Marques2020}); however the shifts are small compared to the overall energy scale of the comparison with theory.\\

\subsection{Sample growth and characterization}
\label{app_growth}
Single crystals of \ce{Sr2RuO4} were grown by a flux feeding floating zone (FFFZ) with Ru self-flux using a commercial image furnace equipped with double elliptical mirrors and two 2.0 kW halogen lamps (NEC Machinery, model SC1-MDH11020). Details of the FFFZ crystal growth are described in detail elsewhere \cite{fittipaldi_crystal_2005,perry_systematic_2004,fittipaldi_floating_2011}. Several techniques, including x-ray diffraction, energy dispersive spectroscopy (EDS) and polarized light optical microscopy (PLOM) analysis, have been used to fully characterize the structure, quality, and purity of the crystals.\\[0.5cm]

\textbf{Data availability:}
The data underpinning the findings of this study
are available online\cite{Kreisel2021}.

\textbf{Code availability:}
The computational data and source code are available upon reasonable request to A. Kreisel (\href{mailto:kreisel@itp.uni-leipzig.de}{kreisel@itp.uni-leipzig.de}).

\textbf{Acknowledgements:}
We acknowledge useful discussions with B.M. Andersen, J.C. Davis, E. Fradkin, S.A. Kivelson, V. Madhavan, A.W. Rost, and A. R{\o}mer.  
P. J. H. was
supported by the U.S. Department of Energy under Grant No. DE-FG02-05ER46236.
CAM and PW acknowledge funding from EPSRC through EP/L015110/1 and EP/R031924/1, respectively, and LCR from the Royal Commission for the Exhibition of 1851.
A portion of this work (T.B. and X.K.) was conducted at the Center For Nanophase Materials Sciences which is a US Department of Energy Office of Science User Facility. This research used resources of the Compute and Data Environment for Science (CADES) at the Oak Ridge National Laboratory, which is supported by the Office of Science of the U.S. Department of Energy under Contract No. DE-AC05-00OR22725.
This manuscript has been authored by UT-Battelle, LLC under Contract No. DE-AC05-00OR22725 with the U.S. Department of Energy. The United States Government retains and the publisher, by accepting the article for publication, acknowledges that the United States Government retains a non-exclusive, paid-up, irrevocable, world-wide license to publish or reproduce the published form of this manuscript, or allow others to do so, for United States Government purposes. The Department of Energy will provide public access to these results of federally sponsored research in accordance with the DOE Public Access Plan (https://www.energy.gov/downloads/doe-public-access-plan).\\

\textbf{Author contributions:}
A.K. and C.A.M. contributed equally to this work.
A.K. performed the calculations with input from T.B. and X.K. who did the ab-initio calculations. C.A.M. and L.C.R. carried out STM experiments. C.A.M. analysed the data. R.F., V.G. and A.V. grew and characterized the samples. A.K., P.W. and P.J.H. wrote the manuscript with key contributions from C.A.M. and L.C.R. All authors discussed and contributed to the manuscript.\\
 
\textbf{Competing interests:} The authors declare no competing interests.\\

\bibliography{bibliograph_SRO_STM.bib}

\ifarXiv
    

\section*{Supplementary material}

# Supplementary Information: Quasiparticle Interference of the van-Hove singularity in Sr$_2$RuO$_4$

A. Kreisel,[1, *] C. A. Marques,[2, *] L. C. Rhodes,[2] X. Kong,[3] T. Berlijn,[3]
R. Fittipaldi,[4] V. Granata,[5] A. Vecchione,[4] P. Wahl,[2] and P. J. Hirschfeld[6]

[1] *Institut für Theoretische Physik, Universität Leipzig, D-04103 Leipzig, Germany*
[2] *SUPA, School of Physics and Astronomy, University of St Andrews,*
*North Haugh, St Andrews, Fife KY16 9SS, United Kingdom*
[3] *Center For Nanophase Materials Sciences, Oak Ridge National Laboratory, Oak Ridge, TN 37831, USA*
[4] *CNR-SPIN, UOS Salerno, Via Giovanni Paolo II 132, Fisciano, I-84084, Italy*
[5] *Dipartimento di Fisica E. R. Caianiello, Universit di Salerno, I-84084 Fisciano, Salerno, Italy*
[6] *Department of Physics, University of Florida, Gainesville, Florida 32611, USA*
(Dated: August 19, 2021)

**This PDF file includes:**

Supplementary Note 1. Band structure details

Supplementary Note 2. Wannier functions

Supplementary Note 3. Staggered on-site order

Supplementary Note 4. Chiral impurity states

Figs. S1 to S4

———————

*These authors contributed equally.

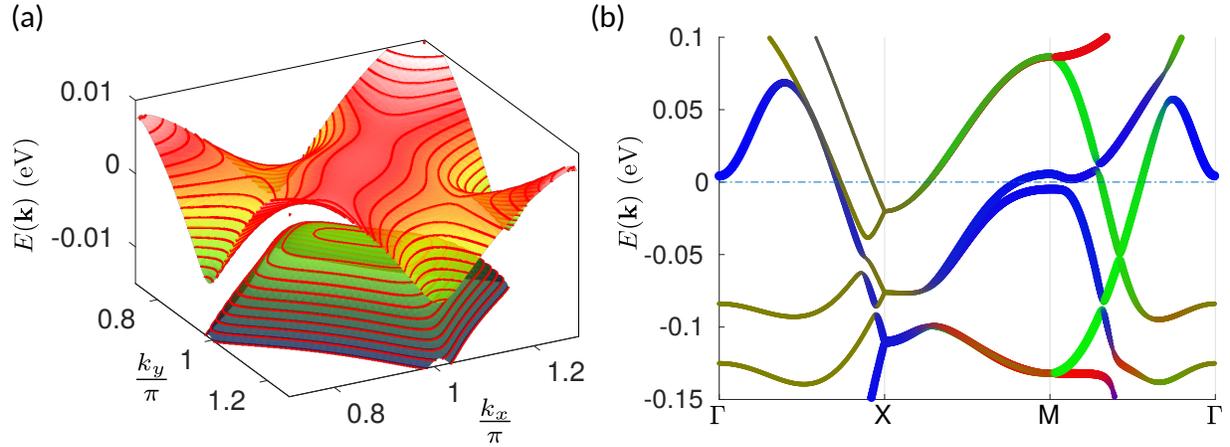

Supplementary Figure 1: **Electronic structure of the reconstructed surface.** (a) Bands close to the M point in the Brillouin zone exhibiting maximum and saddle points obtained from the tight-binding model and leading to the scattering processes from the $\gamma$ band which are most apparent in QPI; red lines are lines of equal energy with spacing of 1 meV. (b) Electronic structure including spin orbit coupling and the nematic reconstruction in a colored fat band scheme (red: $d_{yz}$, green: $d_{xz}$, blue: $d_{xy}$).

## Supplementary Note 1. BAND STRUCTURE FROM TIGHT-BINDING MODEL

In Supplementary Figure 1, we show the band structure obtained form the tight-binding model introduced in the main text close to the M point around the van Hove singularity (Supplementary Figure 1(a)) and a fat-band plot showing the orbital character of the bands crossing the Fermi energy (Supp. Fig. 1(b)).

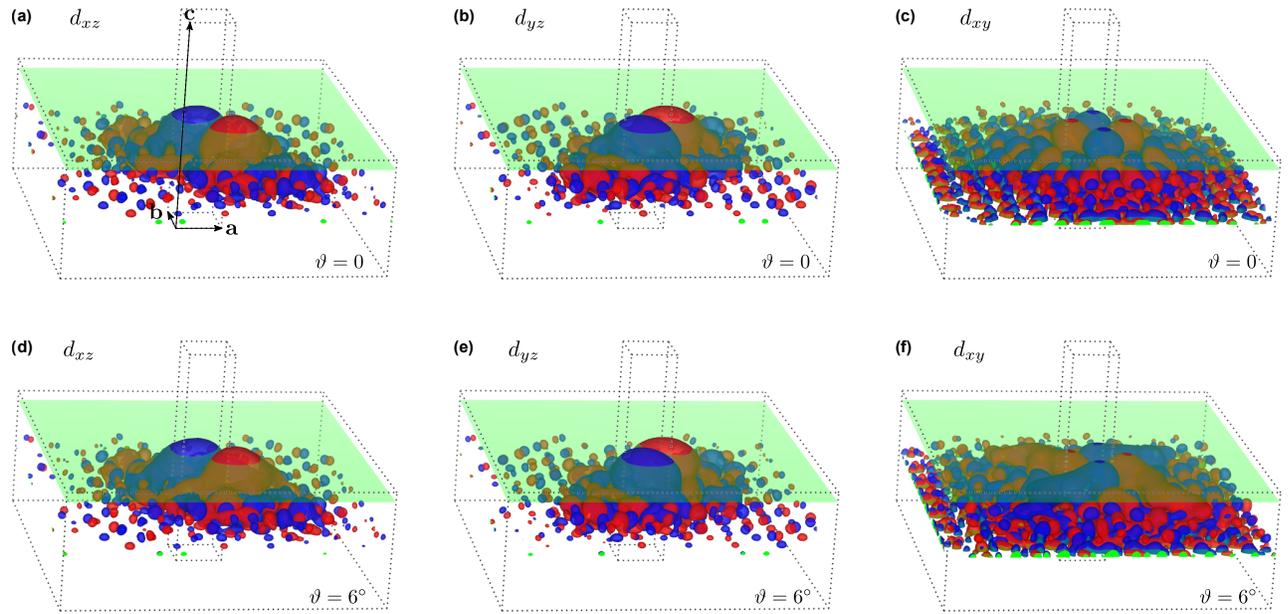

Supplementary Figure 2: **Isosurfaces of Wannier functions in 3D.** (a-c) Wannier functions on the Ru(2) atom together with the elementary cell for the ab-initio calculation with lattice vectors $\mathbf{a}$, $\mathbf{b}$ and $\mathbf{c}$. The Wannier function has been calculated in the large real space box to achieve convergence when summing over the lattice positions $\mathbf{R}$ and $\mathbf{R}'$ in Eq. (D3) of the main text. The green plane is a plane close to $z = 5\text{Å}$ relative to the surface SrO-plane to visualize the dominant contributions for tunneling in terms of the parts of the isosurfaces that stick out. (d-f) Wannier functions for the case including the a $\vartheta = 6°$ rotation of the Ru-O in-plane bonds.

## Supplementary Note 2. WANNIER FUNCTIONS

In Fig. 2 of the main text, we show cuts of the Wannier functions at a representative height above the surface of $Sr_2RuO_4$. To give a more complete picture of the Wannier functions, Supplementary Figure 2 shows isosurface plots of the three Wannier functions of the Ru(2) atom for the case without O octahedron rotation (a-c) and with an O octahedron rotation (d-f) where the strongest changes are visible on the $d_{xy}$ orbital. Note that the corresponding Wannier functions on the Ru(1) atom are just shifted for the case without rotation, and shifted and mirrored to achieve the opposite chirality for the case with octahedral rotation.

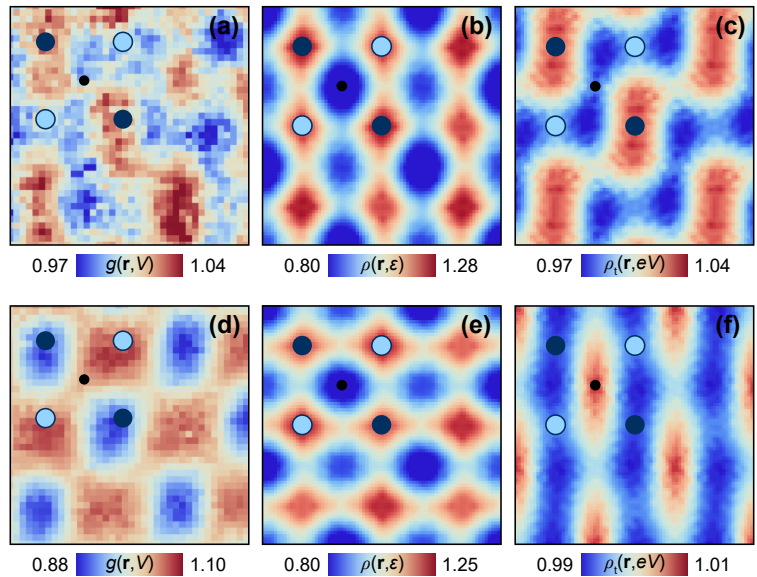

Supplementary Figure 3: **Checkerboard charge order with onsite staggered order.** (a, b) Experimental data shown in Fig. 5 of the main text. (c, d) Continuum LDOS $\rho(\mathbf{r}, E)$ calculated at a height $z = 5\text{Å}$ above the surface at (c) $\epsilon = 5\text{meV}$ and (d) $\epsilon = -5\text{meV}$, but using the staggered order with form factor $f_{\text{bond}}(\mathbf{k}) = 1$. (e, f) Calculated differential conductance map $\rho_t(\mathbf{r}, E)$ at $E = \pm 5\text{meV}$, for $E_{\text{set}} = 10\text{meV}$ for the same model. The intensity of all images has been normalized by the spatial average, color bars indicate relative intensity.

## Supplementary Note 3. STAGGERED ON-SITE ORDER

In this supplementary note, we discuss the effect of different staggered order parameters on the calculated tunneling. Employing the staggered on-site order (see Methods section of the main text), there are few differences in the resulting spectra and maps as we demonstrate in Supplementary Figure 3 which compares the experimental data as presented in the main text (Fig. 4) with the theoretical results for the continuum LDOS $\rho(\mathbf{r}, E)$ and the simulated differential conductance maps $\rho_t(\mathbf{r}, E)$ yielding qualitatively similar results, with minor quantitative differences.

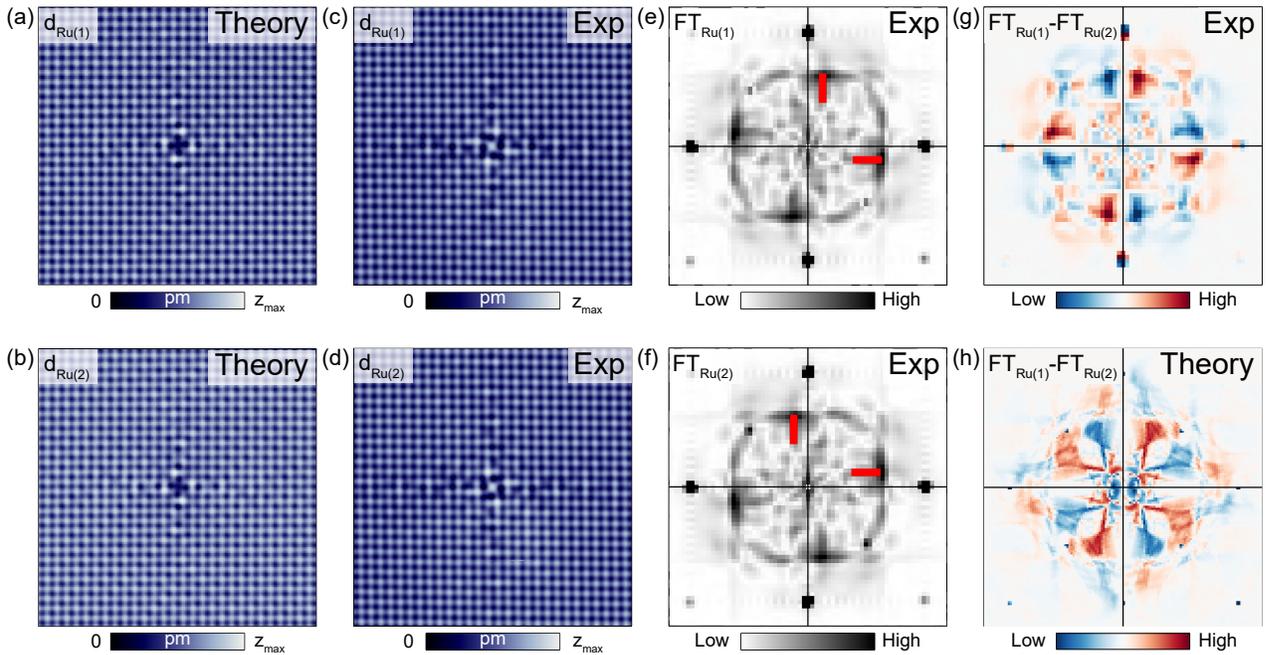

Supplementary Figure 4: **Chiral defects in topographic images.** (a) and (b) Calculated topography with $V_{bias} = 5$ mV close to an individual defect ($V_{imp} = 0.1$eV) in the Ru(1) position and in the Ru(2) position, respectively. ($z_{max} = 34$ pm for (a) and $z_{max} = 37$ pm for (b)). Chiral patterns are seen close to the defects. (c) and (d) Experimental topographies centred at individual defects at the Ru(1) and Ru(2) sites, respectively ($V_{set} = 5$mV, $I_{set} = 100$pA, $T = 500$ mK, $z_{max} = 20$ pm). The defects show two different chiralities due to the presence of the surface reconstruction. Additionally, chiral quasiparticle interference patterns are observed, emanating in the horizontal and vertical direction, in agreement with (a) and (b). (e) and (f) show the amplitude of the Fourier transformation of (c) and (d), respectively. The QPI patterns exhibit chirality dependent on the defect site, with maxima off the high-symmetry directions indicated by the red lines. (g) The difference between (e) and (f), emphasizing the chiral nature of the QPI patterns. (h) Difference between the Fourier transformations of (a) and (b) showing excellent agreement with (g), demonstrating the fidelity of our model in capturing the essential ingredients of the surface electronic structure.

## Supplementary Note 4.  CHIRAL IMPURITY STATES

The rotation of the O octahedra in the ab-initio calculations doubles the elementary cell resulting in a unit cell with two Ru atoms. These reside at the centre of oxygen octahedra which exhibit opposing rotation, clockwise and anti-clockwise direction, through which they acquire opposing chirality and making them distinguishable. This property is seen in experimental data, most notably in topographies as we show in Supplementary Figure 4(c,d) where two defects centered at Ru positions of the two different sublattices are visible. In panel (a,b) of the same figure, we show calculated topographies for the same parameters which show chiral impurity states with rotation direction clockwise and anti-clockwise depending on whether the impurity potential is located on the Ru(1) atom or the Ru(2) atom.


\fi

\end{document}